\documentclass[prl,twocolumn,amssymb,floatfix]{revtex4-1}
\usepackage{amsmath} 
\usepackage{amsfonts} 
\usepackage{amssymb}
\usepackage{bbm}
\usepackage{epsfig}
\usepackage{graphics}
\usepackage{graphicx}
\usepackage{color}

\def\prd{{\sl Phys.\ Rev.\ D \ }}

\newcommand{\lsim}{\,\lower2truept\hbox{${<\atop\hbox{\raise4truept\hbox{$\sim$}}}$}\,}
\newcommand{\gsim}{\,\lower2truept\hbox{${>\atop\hbox{\raise4truept\hbox{$\sim$}}}$}\,}
\newcommand{\pp}{~~~.}
\newcommand{\vv}{~~~,}

\newcommand{\be}{\begin{equation}}
\newcommand{\ee}{\end{equation}}
\newcommand{\bea}{\begin{eqnarray}}
\newcommand{\eea}{\end{eqnarray}}
\newcommand{\beann}{\begin{eqnarray*}}
\newcommand{\eeann}{\end{eqnarray*}}
\newcommand{\benn}{\begin{equation*}}
\newcommand{\eenn}{\end{equation*}}

\newcommand{\at}[2]{\left. #2 \right|_{#1}}

\definecolor{boxcolor}{rgb}{.9,.9,.95}
\definecolor{bordercolor}{rgb}{0.0,0.0,0.4}
\newcommand{\mybox}[1]{ \begin{center}\fcolorbox{bordercolor}{boxcolor}{\parbox{0.475\textwidth}{#1}}\end{center} }

\begin{document}

\title[Clarifying spherical collapse in coupled dark energy cosmologies]
{Clarifying spherical collapse in coupled dark energy cosmologies}

\author{Nico Wintergerst$^{1,3}$, Valeria Pettorino$^{2,3}$}
\affiliation{
$^1$ Arnold-Sommerfeld-Center, Ludwig-Maximilians-Universit\"at, Theresienstr. 37, D-80333 M\"unchen, Germany,\\
$^2$ SISSA, Via Bonomea 265, 34136 Trieste, Italy,
$^3$ Institut  f\"ur Theoretische Physik, Universit\"at Heidelberg,
Philosophenweg 16, D-69120 Heidelberg, Germany.
}

\begin{abstract}
The spherical collapse model is often used to follow the evolution of overdensities into the nonlinear regime. 
We describe the correct approach to be used in coupled dark energy cosmologies, where
a fifth force, different from gravity and mediated by the dark energy scalar field, influences the collapse. 
We reformulate the spherical collapse description by deriving it directly from the set of nonlinear hydrodynamical Navier-Stokes equations. By comparing with the corresponding relativistic equations,
we show how the fifth force should be taken into account within the spherical collapse picture and clarify
the problems arising when an inhomogeneous scalar field is considered within a spherical collapse picture.
We then apply our method to the case of coupled quintessence, where the fifth force acts among cold dark matter particles,
and to growing neutrino quintessence, where the fifth force acts between neutrinos.
Furthermore, we review this method within standard cosmologies and apply our analysis to minimally coupled quintessence. We also check past results for early dark energy parametrizations.
\end{abstract}

\maketitle

\section{Introduction}
A wide variety of theoretical cosmological models can be challenged and discriminated thanks to predictions on 
structure formation. At the nonlinear level the behavior of $\Lambda$CDM cosmologies, in which the role of dark energy (DE) is played by a cosmological constant,
can significantly differ from dynamical dark energy models. In more realistic scenarios, these allow for DE couplings to other species. Interacting dark energy cosmologies include: coupled quintessence
(DE evolution is coupled to dark matter) \cite{amendola_2000, mangano_etal_2003, amendola_quercellini_2003, amendola_2004, wang_etal_2007, 
pettorino_baccigalupi_2008, quartin_etal_2008, boehmer_etal_2008, bean_etal_2008, lavacca_etal_2009}; growing neutrino cosmologies \cite{amendola_etal_2007, wetterich_2007, mota_etal_2008}
 and MaVaNs (\cite{fardon_etal_2004, afshordi_etal_2005, bjaelde_etal_2007, brookfield_etal_2006, brookfield_etal_2007, takahashi_tanimoto_2006} and references therein) (DE is interacting with neutrinos); 
so-called modified gravity theories such as scalar-tensor theories, including F(R) and extended quintessence
\cite{hwang_1990, hwang_1990_2, wetterich_1995, uzan_1999, faraoni_2000, riazuelo_uzan_2002,               
perrotta_etal_2000, boisseau_etal_2000, perrotta_baccigalupi_2002, pettorino_etal_2005}. In all these cosmologies a fifth force is present, acting on species whose evolution is coupled to the 
DE evolution. The presence of a fifth force, mediated by the DE scalar field (the cosmon, seen as the mediator of a cosmological 
interaction) can modify structure formation in a significant way \cite{baldi_etal_2009, maccio_etal_2003}, in particular at large scales 
\cite{wintergerst_etal_2009}.
In view of future data, it is therefore important to understand how these theories behave when density perturbations reach 
nonlinearity.
  
While up to now N-body simulations represent the best way to numerically evolve structures, 
other semi-analytical methods have been used to follow perturbations into the nonlinear regime, either using spherical collapse 
\cite{peebles_1967, gunn_gott_1972, padhmanabhan, peacock, bilic_etal_2003, pace_etal_2010} or other alternative methods \cite{peacock_dodds_1996, pietroni_2008, angrick_bartelmann_2010}. 
In particular, spherical collapse has been used in several occasions in literature
 for $\Lambda$CDM \cite{padhmanabhan, peacock, engineer_etal_1998}, minimally coupled quintessence models \cite{wang_steinhardt_1998, mainini_etal_2003, mota_vandebruck_2004, maor_lahav_2005, wang_2005, dutta_maor_2006, mota_etal_2007, abramo_etal_2007, creminelli_etal_2009}, coupled quintessence
 \cite{mainini_bonometto_2006, nunes_mota_2006} and when parametrizing early dark energy 
contributions \cite{bartelmann_etal_2006, sadeh_etal_2007, francis_etal_2008}. 

In this paper we give a detailed description of the spherical collapse method and clarify some tricky issues in its applications.
We lay particular focus on the calculation of the extrapolated linear density contrast 
at collapse $\delta_c$, a quantity of major interest within a spherical collapse description, often used in a Press-Schechter \cite{press_schechter_1974} approach to estimate dark matter halo mass distributions.

After reviewing results for standard cosmologies like $\Lambda$CDM, 
we consider the case in which a fifth force is present in addition 
to standard gravitational attraction, as in the case of all the interacting dark energy models mentioned above. 
The inclusion of the fifth force within the spherical collapse picture requires particular attention.
Spherical collapse is intrinsically based on gravitational attraction only and cannot account
 for other external forces unless it is suitably modified. The dynamics in the spherical collapse model are governed by 
Friedmann equations. Hence, only gravitational forces determine the evolution of the different scale factors and, in turn, of the density contrast. 

A detailed comparison between the linearized spherical collapse picture and the linear relativistic equations allows us to 
first identify the presence (or absence), in the spherical collapse picture, of terms which are
a direct signature of the coupling already at the linear level. 
We show how spherical collapse necessitates to be suitably modified whenever an additional force other
than gravity is present and is big enough to influence the collapse.
We use this comparison also to show that a standard treatment of spherical collapse may lead to problems even in the uncoupled case when treating inhomogeneities in the scalar field. 

A modification of the spherical collapse picture is indeed possible, via a nonlinear analysis of the model. 
We derive the set of second order differential equations for the density contrast from the nonlinear Navier-Stokes equations described in \cite{wintergerst_etal_2009}, extending an idea from \cite{wintergerst_2009}. We show how $\delta_c$ can be evaluated directly from these equations and how they can serve as a starting point for a reformulation of spherical collapse. Our results match the numerical solution
of the nonlinear hydrodynamical equations solved as described in \cite{wintergerst_etal_2009, wintergerst_2009}.

We apply our method to coupled quintessence scenarios where a coupling is present 
among dark matter particles, comparing our results with alternative methods presented
 in the past \cite{mainini_bonometto_2006, nunes_mota_2006}.
As a further application of our method, we consider for the first time spherical collapse within growing neutrino models,
 where an interaction is active among neutrinos: in this case we
obtain an extrapolated linear density at collapse $\delta_c$ which shows an oscillating behavior, a characteristic
feature of the interaction.

Finally we confirm results found in \cite{francis_etal_2008} on spherical collapse and early dark energy (EDE).

In Sec.~\ref{sec:spher_coll} we recall the spherical collapse model and its applications to standard cosmologies (Sec.~\ref{standard_cosm}). 
In Sec.~\ref{sec:coupled_quint} we focus on spherical collapse in presence of a fifth force, 
taking the case of coupled quintessence as an example of fifth force cosmologies.
In Sec.~\ref{sec:spher_coll_coupled_inc}, we demonstrate that the standard spherical collapse leads to wrong results 
when applied to coupled quintessence: indeed, by comparing with full relativistic equations (Sec.\ref{relativistic_eqts}) 
we demontrate that the spherical collapse equations lack terms that are essential in the presence of a fifth force (Sec.\ref{lack_fifthforce}) and
can lead to incorrect results when an inhomogeneous DE scalar field is included within this framework (Sec. \ref{inhomogeneity}).
Consequently, in Sec.~\ref{sec:spher_coll_coupled_cor}, we illustrate how the spherical collapse can be correctly reformulated in coupled scenarios by basing it on the full nonlinear Navier-Stokes equations for the respective model. We further comment on the careful choice of initial conditions in Sec.~\ref{sec:ini_conds}. We apply the derived formalism to give results for coupled quintessence (Sec.~\ref{sec:coupled_results}) and growing neutrinos (Sec.~\ref{sec:nu_results}). Finally, we use the described framework to confirm results found in \cite{francis_etal_2008} for uncoupled early dark energy.

\section{Spherical Collapse}\label{sec:spher_coll}

Consider a cold dark matter density perturbation within a homogeneous background Universe. Under the effect of gravitational attraction the perturbation grows, possibly entering the nonlinear regime, depending on the scale of the perturbation. A popular method often used to follow the evolution of cold dark matter (CDM) structures during the first stages of the nonlinear regime is the spherical collapse model. In its original applications \cite{peebles_1967, gunn_gott_1972, peacock, padhmanabhan}, it is assumed that the initial overdensity obeys a top hat profile \be \delta\rho_{\text{in}} (t,s) \equiv \delta\rho_0(t) \Theta(r(t) - s) \vv \ee where $r(t)$ specifies the radius of the top hat and $s$ is the spherical coordinate indicating the distance from the center of the perturbation. $\Theta(r(t) - s)$ is the top hat function, equal to $1$ for $s \le r(t)$ and $0$ otherwise.
 The amplitude of the top hat is given by $\delta \rho_0$ and is evolving in time.
As a consequence of Birkhoff's theorem of General Relativity, which ensures that the dynamics of the radius $r(t)$ are governed only by the enclosed mass, the top hat ``bubble'' is
 conveniently described as a closed Universe where the total density $\rho = \rho_{crit} + \delta \rho_{m}$ exceeds the critical density
$\rho_{crit}$ due to the presence of
 CDM density perturbation. 

Hence, all densities and geometric quantities are treated according to the Friedmann equations:
\bea 
\label{eq:sc_f1b} H^2 &\equiv& \left(\frac{\dot r}{r}\right)^2 = \frac{1}{3} \sum_{\alpha} \rho_{\alpha} - \frac{K}{r^2} \vv \\
\label{eq:sc_f2b} \frac{\ddot r}{r} &=&  - \frac{1}{6} \sum_{\alpha} \left[\rho_{\alpha}(1 + 3 w_{\alpha})\right] \pp
\eea
Here the ``scale factor'' is given by the radius of the bubble $r(t)$, commonly normalized to match the background scale factor $a(t_i)$ at some initial time $t_i$. The corresponding Hubble function of the bubble
 is indicated by $H$. Eq.\,(\ref{eq:sc_f1b}) explicitly contains a curvature term $K$; Eq.\,(\ref{eq:sc_f2b}), albeit the lack of an explicit curvature term, is still describing a closed Universe as the sum of the densities on the right hand side exceeds the critical one. Note that throughout this work densities have been normalized in units of the square of the reduced Planck mass $M^2 = \left(8\pi G_N\right)^{-1}$. 

The bubble is embedded in a homogeneous Friedmann-Robertson-Walker (FRW) background characterized by a scale factor $a(t)$ and a corresponding Hubble
 function $\bar H \equiv {\dot a}/a$. We use a bar to indicate background quantities. For clarity,
 we recall the Friedmann equations describing the homogeneous and flat background Universe:

\bea 
\label{eq:sc_f1bkg} {\bar H}^2 &\equiv& \left(\frac{\dot a}{a}\right)^2 = \frac{1}{3} \sum_{\alpha} \bar{\rho}_{\alpha} \vv \\
\label{eq:sc_f2bkg} \frac{\ddot a}{a} &=& - \frac{1}{6} \sum_{\alpha} \left[\bar{\rho}_{\alpha}(1 + 3 \bar{w}_{\alpha})\right] \pp
\eea

Note that throughout this work we neglect baryonic components. For simpler notation we refer to CDM by a subscript $m$. 

\subsection{Applications to standard cosmologies} \label{standard_cosm}
Spherical collapse can be safely applied to the case of Einstein de Sitter (EdS) cosmologies (in which $\Omega_m = 1$), and to $\Lambda$CDM models.
In this case, the energy density of matter $\rho_m$, appearing on the right hand side of 
Eq.\,(\ref{eq:sc_f2b}) and (\ref{eq:sc_f2bkg}), is conserved both inside and outside the overdensity:
 \bea \label{dens_bub} {\dot\rho}_{m} &=& - 3H (1 + {w}_m) {\rho}_{m} \vv \\ \label{dens_bkg} \dot{\bar\rho}_m &=& -  3{\bar H} (1 + \bar{w}_m) \bar{\rho}_m \pp \eea
The nonlinear density contrast is defined by $1+\delta_m \equiv \rho_m/\bar\rho_m$ and is determined by the above equations.
The linear density contrast evolves according to well known
 linear perturbation theory \cite{kodama_sasaki_1984, ma_bertschinger_1995} and satisfies the linear equation:
 \be \label{eq:lin_delta_uc} \ddot\delta_{m,L} + 2{H} \dot\delta_{m,L} - \frac{3}{2} {H}^2 \Omega_{m} \delta_{m,L} = 0 \vv \ee

\begin{figure}[ht]
\begin{center}
\includegraphics[width=85mm,angle=0.]{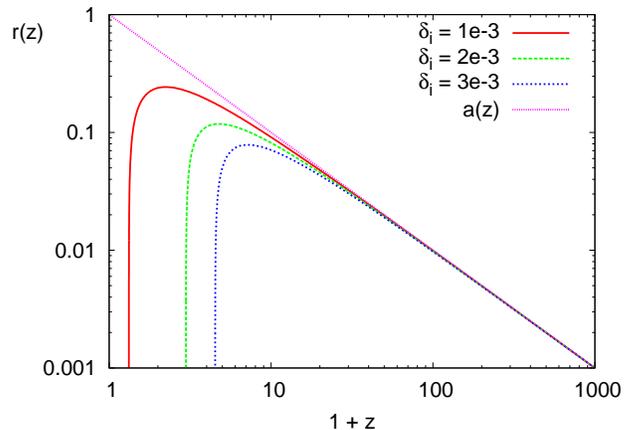}
\end{center}
\caption{Evolution of radial parameter $r(z)$ for different initial overdensities in a $\Lambda$CDM model with $\Omega_m = 0.3$, $\Omega_\Lambda = 0.7$, as used throughout this work. For comparison, we have included the background scale factor $a$ (dotted, pink).}
\label{fig:r_of_z_lcdm}
\vspace{0.5cm}
\end{figure}

Equations (\ref{eq:sc_f2b}) - (\ref{eq:lin_delta_uc}) can be integrated numerically. We start the integration at some initial time $t_\text{in}$ in which
 the total energy density in the bubble is higher than the critical energy density, due to the presence of the CDM 
overdensity $\delta_{m}$. Equation (\ref{eq:sc_f2b}) provides $r(z)$, which is shown
in Fig.{\ref{fig:r_of_z_lcdm}} for a $\Lambda$CDM model with $\Omega_\Lambda = 0.7$ and for three different overdensities ($\delta_\text{in} = 1\cdot10^{-3}, 2\cdot10^{-3}, 3\cdot10^{-3}$ for $z_\text{in} = 10^4$): $r(z)$ first increases as the bubble
 expands with the background; then, it reaches a maximum value (turnaround) in which comoving velocities become zero; finally,
 the bubble collapses, the radius tends to zero and the nonlinear density contrast $\delta_{m}$ increases rapidly. The redshift of collapse depends on the amplitude of the initial perturbation. The higher this is, the earlier the overdense region will collapse. The corresponding
 value of the linear density contrast extrapolated at the time of collapse is usually referred to as $\delta_c$ and represents
 one of the key ingredients for a Press-Schechter analysis, which gives statistical estimates of the cluster distribution in space.
 We will not go into the Press-Schechter procedure here; instead, we will focus on the calculation of $\delta_c$ within the spherical
 collapse analysis, putting in evidence how this calculation has to be carefully performed depending on the underlying theoretical model.

\begin{figure}[ht]
\begin{center}
\includegraphics[width=85mm,angle=0.]{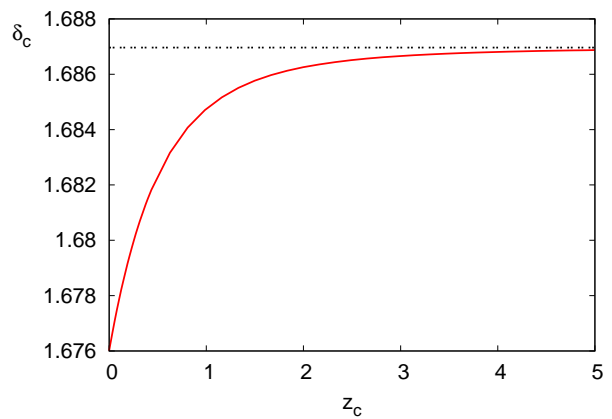}
\end{center}
\caption{Extrapolated linear density contrast at collapse $\delta_c$ vs. redshift at collapse $z_c$ for a $\Lambda$CDM
 (solid, red) and an EdS (double-dashed, black) model.}
\label{fig:delta_c_lcdm}
\vspace{0.5cm}
\end{figure}

In an Einstein de Sitter model the linear density contrast at collapse can be calculated analytically \cite{padhmanabhan, peacock}:
 it is equal to a constant value independent of the redshift of collapse $z_c$ 
\be \label{eq:delta_c_eds} \delta_{c} = (3/20)\left(12\pi\right)^{2/3} \simeq 1.686 \pp \ee
Note that we define $z_c$ as the redshift at which $r \rightarrow 0$.
In a $\Lambda$CDM model one expects this value to decrease for late collapse times \cite{padhmanabhan, peacock}, when dark energy 
dominates over matter and leads to cosmic acceleration, slowing down structure formation. 
In Fig. \ref{fig:delta_c_lcdm} we plot $\delta_c(z_c)$ for both EdS and $\Lambda$CDM.

\begin{figure}[ht]
\begin{center}
\includegraphics[width=85mm,angle=0.]{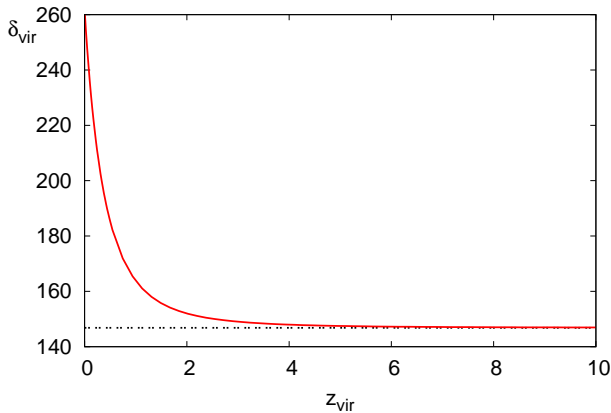}
\end{center}
\caption{Nonlinear density contrast at virialization for a $\Lambda$CDM (solid, red) and an EdS (double-dashed, black) model.}
\label{fig:delta_vir_of_z}
\end{figure}

It is also common to analyze the (nonlinear) density contrast at virialization. 
From the virial theorem one may deduce \cite{padhmanabhan, peacock, engineer_etal_1998, maor_lahav_2005} that a given bubble virializes whenever
 it has collapsed to half its turnaround radius. In an EdS Universe, the density contrast at virialization
 is analytically found to be $\delta_\text{vir} = \left(9\pi + 6\right)^2/8 \simeq 146.8$. 
For the $\Lambda$CDM model, an increase is observed for late collapse times. This corresponds to 
the fact that, in presence of dark energy, it takes longer for structures to virialize, with a corresponding
higher value of $\delta_\text{vir}$, as shown in Fig.\ref{fig:delta_vir_of_z}.

\begin{figure}[ht]
\begin{center}
\includegraphics[width=85mm,angle=0.]{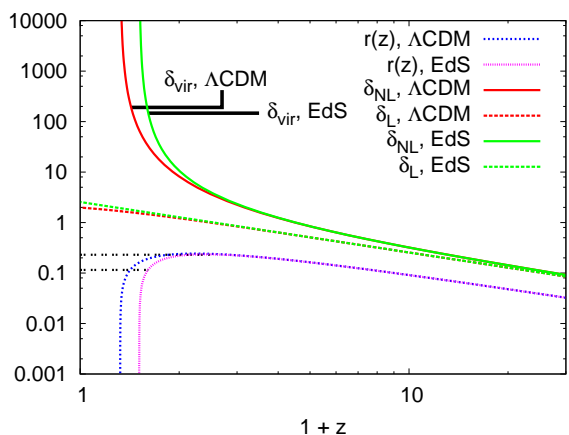}
\end{center}
\caption{Linear and nonlinear density contrasts in $\Lambda$CDM (red) and EdS (green) models, as well as the corresponding radius functions (blue, short-dashed and pink, dotted, respectively). The upper double-dashed black line marks the turn around radius $r_{\text{ta}}$, the lower one $r_{\text{ta}}/2$. The later virialization time in $\Lambda$CDM ($r_{\text{ta}}/2$ is reached substantially later, when the blue dotted line and the lower black double-dashed lines intersect) leads to an increase of $\delta_{\text{vir}}$. The overdensities collapse when the radii go to zero and $\delta_c$ is given by the value reached by the linear curves at this redshift. Although collapse happens later for $\Lambda$CDM, the linear growth is suppressed at late times (red dashed line in comparison to green dashed line). This leads to a decrease in $\delta_c$. }
\label{fig:lcdm_eds_comp}
\end{figure}

To better illustrate the effect of $\Lambda$ on both $\delta_\text{vir}$ and $\delta_c$, we have plotted $\delta_{m,L}$ and $\delta_{m,NL}$
 in Fig.\ref{fig:lcdm_eds_comp}, as well as the radius $r$ for a fixed initial overdensity in both an EdS and a $\Lambda$CDM Universe. It can be seen that the later virialization in $\Lambda$CDM leads to an increase in $\delta_{vir}$. On the other hand, the smaller linear growth rate reduces the extrapolated linear density contrast $\delta_c$ in $\Lambda$CDM.

Alternatively to a cosmological constant, dark energy can be described by a dynamical energy component, such as a quintessence
 scalar field rolling down a potential \cite{wetterich_1988, ratra_peebles_1988}. A meaningful quintessence model should 
naturally explain why dark energy dominates over cold dark matter only at recent times; this happens to be difficult
to achieve within minimally coupled quintessence models, which are often fine-tuned as much as a $\Lambda$CDM model \cite{matarrese_etal_2004}. Viable models in this direction often
 involve the presence of a coupling between the dark energy scalar field, referred to as ``cosmon'' or ``quintessence'', and other
 components in the Universe such as cold dark matter \cite{wetterich_1995, amendola_2000} or neutrinos
 \cite{fardon_etal_2004, amendola_etal_2007, wetterich_2007, mota_etal_2008, wintergerst_etal_2009}. The presence of
 an interaction that couples the cosmon dynamics to another species introduces a new force.
 This ``fifth force'' is acting
 between particles (CDM or neutrinos in the examples mentioned) and is mediated by dark energy fluctuations. Whenever
 such a coupling is existent, spherical collapse, whose concept is based on gravitational attraction, has to be
 suitably modified. In the following sections we will present some examples of quintessence models in presence of a fifth force
and show how the latter can be taken into account.

\section{Coupled quintessence cosmologies}\label{sec:coupled_quint}

The first set of cosmologies in presence of a fifth force that we consider is coupled quintessence (CQ):
here the evolution of the quintessence scalar field (from hereon we refer to it as the ``cosmon'' \cite{wetterich_1988}) is coupled to CDM
 \cite{wetterich_1995, amendola_2000, amendola_2004, pettorino_baccigalupi_2008, baldi_etal_2009}. 
The cosmon ${\bar\phi}$ interacts with CDM particles whose mass $m({\bar\phi})$ changes with ${\bar\phi}$. This set of cosmologies is described by the Lagrangian:
\be \label{L_phi} {\cal L} =
-\frac{1}{2}\partial^\mu {\bar\phi} \partial_\mu {\bar\phi} - U({\bar\phi}) -
m({\bar\phi})\bar{\psi}\psi + {\cal L}_{\rm kin}[\psi] \,, \ee in which the mass of
matter fields $\psi$ coupled to DE is a function of the scalar field ${\bar\phi}
$.

The homogeneous flat background follows the set of equations described in \cite{amendola_2000,mota_etal_2008}.
The Universe evolves in time according to the Friedmann and acceleration equations:
\be \label{f1} {\bar H}^2 \equiv \left(\frac{\dot a}{a}\right)^2 = \frac{1}{3} \sum_\alpha {\bar\rho}_\alpha \ee 
and
\be \label{f2} \frac{\ddot a}{a} = - \frac{1}{6} \sum_\alpha \left[{\bar\rho}_\alpha(1 + 3 {\bar w}_\alpha)\right] \ee
where the sum is taken over all components $\alpha$ in the Universe.
A crucial ingredient is the dependence of CDM mass on
the cosmon field ${\bar\phi}$, as encoded in the dimensionless cosmon-CDM
coupling $\beta$, \be \beta \equiv - \frac{d \ln{m}}{d {\bar\phi}} \pp
\ee For increasing ${\bar\phi}$ and $\beta > 0$ the mass of CDM particles decreases with
time \be m = \bar{m} e^{-{{\beta}} {\bar\phi}} \vv \ee where
$\bar{m}$ is a constant and $\beta$ is also fixed to be a constant in the simplest coupling case. The cosmon field ${\bar\phi}$ is normalized in units of the reduced Planck mass $M = (8 \pi G_N)^{-1/2}$, and $\beta \sim 1$ corresponds to a cosmon-mediated interaction for CDM particles of roughly gravitational strength. 

For a given cosmological model with a set time dependence of ${\bar\phi}$, one can determine
the time evolution of the mass $m(t)$. The dynamics of the cosmon can be inferred from the Klein
Gordon equation, now including an extra source due to the coupling to CDM:

\be \label{kg} \ddot{\bar\phi} + 3{\bar H} \dot{\bar\phi} + \frac{dU}{d {\bar\phi}} = \beta {\bar\rho}_m \,\, \pp \ee
We choose an exponential potential \cite{wetterich_1988, ratra_peebles_1988,  ferreira_joyce_1998, barreiro_etal_2000}:

\be \label{pot_def} V({\bar\phi}) = M^2 U({\bar\phi}) = M^4 e^{- \alpha {\bar\phi}} \vv \ee
where the constant $\alpha$ is one of the free parameters of our model.
Note that our analysis, however, is more general and can be applied in presence of any quintessence potential.

The homogeneous energy density and pressure of the scalar field ${\bar\phi}$ are defined
in the usual way as \be \label{phi_bkg} {\bar\rho}_{\phi} = \frac{\dot{\bar\phi}^2}{2} + U({\bar\phi})  \vv \,\,\, {\bar p}_{\phi} = \frac{\dot{\bar\phi}^2}{2} - U({\bar\phi})  \vv \,\,\, {\bar w}_{\phi} = \frac{{\bar p}_{\phi}}{{\bar\rho}_{\phi}} \pp \ee
Finally, we can express the conservation equations for dark energy and coupled
matter as follows \cite{wetterich_1995, amendola_2000}: 
\bea \label{cons_phi} \dot{\bar\rho}_{\phi} &=& -3 {\bar H} (1 + {\bar w}_\phi) {\bar\rho}_{\phi} +
\beta \dot{\bar\phi} {\bar\rho}_m \vv \nonumber \\
\label{cons_gr} \dot{\bar\rho}_m &=& -3 {\bar H} {\bar\rho}_m - \beta \dot{\bar\phi}  {\bar\rho}_m
\pp \eea The sum of the energy momentum tensors for CDM and the cosmon is conserved,
 but not the separate parts. We neglect a possible cosmon coupling to baryons ($b$) or
 neutrinos ($\nu$), so that $\label{cons_alpha} \dot{\bar\rho}_{b, \nu} = -3 {\bar H}(1 + {\bar w}_{b, \nu}) {\bar\rho}_{b, \nu} $.

\begin{figure}[ht]
\begin{center}
\includegraphics[width=85mm,angle=0.]{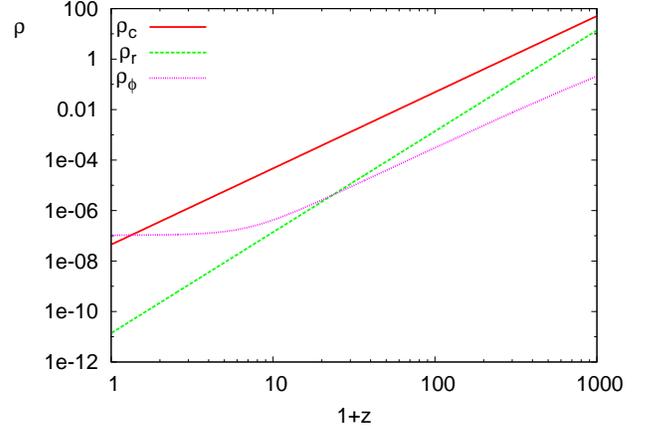}
\end{center}
\caption{Energy densities of cold dark matter (solid), dark energy (dotted) and radiation (long dashed) are plotted vs redshift. We take a constant $\beta = 0.1$, with $\alpha = 0.1$.}
\label{fig_1}
\vspace{0.5cm}
\end{figure}

For a given potential \eqref{pot_def} the evolution equations for the different species can be numerically
 integrated, giving the background
evolution shown in Fig.\ref{fig_1} (for constant $\beta$). For a detailed description of attractor
 solutions in this context see \cite{wetterich_1995, copeland_etal_1998, amendola_2000}

It will later prove useful to understand the evolution of perturbations within coupled models in the linear regime.
The relativistic calculation in coupled quintessence is described in detail in \cite{amendola_2004, mota_etal_2008, wintergerst_etal_2009}. 
Here we just recall the resulting second order equations (in Fourier space) for $\delta_{m,L}$ and the perturbation of the scalar field $\delta \phi$ in Newtonian gauge (in which the nondiagonal metric perturbations are fixed to zero):

\bea \ddot\delta_{m,L} &=& -2\bar{\bar H} \left(\dot\delta_{m,L} + \beta\,\delta\dot\phi\right) + (k/a)^2\left(\Phi + \beta\,\delta\phi\right) \nonumber \\ 
&& + \beta\,\dot{\bar\phi}\left(\dot\delta_{m,L} + \beta\,\delta\dot\phi\right) - 3\dot\Phi\left(2\bar{\bar H} - \beta\,\dot{\bar\phi}\right) \nonumber \\
 \label{eq:delta_lin_nonnewt}  && - 3\ddot\Phi - \beta\,\delta\ddot\phi \vv \eea 
  \bea \delta\ddot\phi &=& -3 \bar{\bar H}\delta\dot\phi - U_{,\phi\phi}\,\delta\phi + \beta{\bar\rho}_m\left(\delta_{m,L} + 2\Phi\right) \nonumber \\
\label{eq:phi_lin_nonnewt} && - (k/a)^2\delta\phi - 2\Phi U_{,\phi\phi} + 4\dot{\bar\phi}\dot\Phi \pp \eea

Since the spherical collapse is intended to model the nonlinear evolution in the Newtonian limit, we are interested in the case in which $k \gg a\bar H$.
Following \cite{amendola_2004, pettorino_baccigalupi_2008}, we obtain
\be \label{deltadotdot_lin} \ddot\delta_m = (\beta \dot{\bar\phi} - 2\bar H) \dot\delta_m + (k/a)^2 (1+2 \beta^2)\Phi \vv \ee
\be
\label{eq:deltaphi_newt_limit} k^2 \delta \phi \sim \beta\,a^2 {\bar\rho}_m \delta_m \pp
\ee and 
the gravitational potential is approximately given by
\be 
\label{eq:grav_pot_newt_limit} k^2 {\Phi} \sim \frac{1}{2} a^2 \sum_{\alpha \neq \phi} {\bar\rho}_\alpha \delta_\alpha \, \vv  \ee where we have assumed that no anisotropic stress is present, so that $\Phi = -\Psi$. 
We can then
define an effective gravitational potential as 
\be \label{eq:eff_grav_pot} {\Phi_{\text{eff}}} \equiv {\Phi} + {\beta} \delta \phi \pp \ee 
In real space (comoving spatial coordinates) and after substituting the expressions
 for ${\Phi}$ [Eq.~(\ref{eq:grav_pot_newt_limit})] and for $\delta \phi$ [Eq.~(\ref{eq:deltaphi_newt_limit})], we get the modified Poisson equation: 
\be \label{eq:laplace_grav_pot} \Delta {\Phi_{\text{eff}}} = -\frac{a^2}{2} {\bar\rho}_m \delta_m \left(1+2 \beta^2\right) \, \pp \ee
Cold dark matter then feels an effective gravitational constant 
\be \label{eq:G_eff} \tilde{G}_{\text{eff}} = G_{N}[1+2\beta^2] \vv \ee
where $G_{N}$ is the usual Newton's constant.

The first term on the right hand side of Eq.(\ref{deltadotdot_lin}) includes the expansion damping, modified by the velocity dependent term $\beta \dot{{\bar\phi}}$, which accounts for momentum 
conservation; the last term on the right hand side specifies the presence of the fifth force.

\section{Standard spherical collapse and coupled quintessence}\label{sec:spher_coll_coupled_inc}

We will now apply the framework described in Sec. \ref{sec:spher_coll} as it is to CQ. This
approach has been used, for example, in \cite{nunes_mota_2006}. We will show, by comparison with the perturbation equations recalled
 in Sec.\ref{sec:coupled_quint}, that this does not correctly model the evolution of nonlinear structures in coupled quintessence.

For this purpose, consider the standard spherical collapse equations (\ref{eq:sc_f1b}) and (\ref{eq:sc_f2b}), in which the densities on the right hand side satisfy the coupled conservation equations:
\bea
\label{rad_bub} \dot\rho_{r} &=& -4 {H} \rho_{r} + \Gamma_r \vv \\
\label{mat_bub} \dot\rho_{m} &=& -3 {H} \rho_{m} - \beta \dot\phi  \rho_{m} \vv \\ 
\label{rhophi_bub} \dot\rho_{\phi} &=& -3 {H} (\rho_{\phi} + p_{\phi}) + \beta \dot\phi  \rho_{cdm} + \Gamma_{\phi} \vv \eea
or equivalently the Klein-Gordon equation:
\be \label{kg_bub} \ddot\phi + 3H\,\dot\phi + U_{,\phi} = \beta \rho_m + \frac{\Gamma_{\phi}}{\dot\phi} \pp \ee
Here additional source terms $\Gamma_r$ and $\Gamma_\phi$ may account for possible differences between the bubble and the background
 components for radiation and the scalar field, respectively \cite{mota_vandebruck_2004, maor_lahav_2005, wang_2005, nunes_mota_2006}. In case of clustering dark energy and/or radiation, both source terms are set to zero:
\bea 
\label{rad_source_inhom} \Gamma_r &\equiv& 0 \vv \\
\label{phi_source_inhom} \Gamma_{\phi} &\equiv& 0 \pp
\eea
If both radiation and the scalar field are to be homogenous, i.e. they behave in the bubble as in the background, the source terms are defined as:
\bea 
\label{rad_source} \Gamma_r &\equiv& 4({H} - \bar{H})\rho_{r} \vv \\
\label{phi_source} \Gamma_{\phi} &\equiv& 3({H} - \bar{H})\,(\rho_{\phi} - p_{\phi}) + \beta\,(\dot\phi - \dot{\bar\phi}) \pp
\eea
In order to account for a fractional outflow of dark energy or radiation, one may suitably interpolate between the two values \cite{mota_vandebruck_2004,maor_lahav_2005,wang_2005,nunes_mota_2006}.

\subsection{Comparison with relativistic equations} \label{relativistic_eqts}

We will now show that the approach described by Eqs. (\ref{eq:sc_f2b}) and (\ref{eq:sc_f2bkg}) together with
 (\ref{rad_bub} - \ref{phi_source}) is incorrect. The actual fifth force term is entirely missing from the equations. 

As a starting point, we remark that Eq.(\ref{cons_gr}) for the background CDM density
 and (\ref{mat_bub}) for the bubble CDM density can be directly integrated \cite{amendola_2000} to yield
\bea \label{eq:cdm_back} \bar\rho_{m} &=& \bar\rho_{m,\text{in}}\,e^{\beta\,\bar\phi_\text{in}}\,\left(\frac{a}{a_\text{in}}\right)^{-3}\,e^{-\beta\,\bar\phi} \vv \\
\label{eq:cdm_bub} \rho_{m} &=& \rho_{m,\text{in}}\,e^{\beta\,\phi_\text{in}}\,\left(\frac{r}{r_\text{in}}\right)^{-3}\,e^{-\beta\,\phi} \pp \eea
The density contrast $\delta_m$ is then given by
\bea \label{eq:sc_dens_cont_temp} 1 &+& \delta_{m} \equiv \frac{\rho_{m}}{\bar\rho_{m}} \nonumber \\
&=& (1 + \delta_{m,\text{in}})\,e^{\beta\,\delta\phi_\text{in}}\,\left(\frac{r_\text{in}}{a_\text{in}}\right)^3\left(\frac{a}{r}\right)^3\,e^{-\beta\,\delta\phi} \vv \eea
where we have introduced $\delta\phi \equiv \phi - \bar\phi$. 
The first and second time derivatives of $\delta_m$ read
\bea \label{eq:sc_dens_cont_dot} \dot\delta_m &=& 3 \left(1 + \delta_m\right) \left(\bar{H} - {H}\right) - \beta\,\delta\dot\phi \left(1 + \delta_m\right) \vv \\
\ddot\delta_m &=& \frac{\dot\delta_m^2}{1 + \delta_m} + 3\left(1 + \delta_m\right)\,\left( \dot{\bar H} - {\dot H} \right) \nonumber \\
\label{eq:sc_dens_cont_ddot} && - \beta\,\delta\ddot\phi\,\left(1 + \delta_m\right) \pp \eea
We can substitute $\left(\bar{H} - {H}\right)$ and $\left( \dot{\bar H} - {\dot H} \right)$ using Eqs.\,(\ref{eq:sc_f2bkg}) and (\ref{eq:sc_f2b}).
Taking the square of Eq.\,(\ref{eq:sc_dens_cont_dot}) and inserting it into Eq.\, (\ref{eq:sc_dens_cont_ddot})
we obtain: 
\bea \ddot\delta_m &=& -2\bar{H}\left(\dot\delta_m + \beta\,\delta\dot\phi\left(1+\delta_m\right)\right) \nonumber \\
&+& \frac{1}{2}\,\left(1+\delta_m\right)\sum_{\alpha}\left(\delta\rho_{\alpha} + 3\delta{p}_{\alpha}\right) + \frac{4}{3}\,\frac{ {\dot\delta_m}^2 }{1 + \delta_m} \nonumber\\
&+& \frac{2}{3}\beta\,\delta\dot\phi\,\dot\delta_m + \frac{1}{3}\,(1 + \delta_m)\,\beta^2\,{\delta\dot\phi}^2 \nonumber \\
\label{eq:sc_dens_cont_evol} &-& \beta\,\delta\ddot\phi\left(1 + \delta_m\right) \eea 
This is the evolution equation for the density contrast $\delta_m$, as derived directly from spherical collapse applied to coupled quintessence.
Usually, one considers only cold components to actually cluster, reducing the sum in Eq.\,(\ref{eq:sc_dens_cont_evol}) to one over CDM only. For the moment we still allow for an inhomogenous scalar field and therefore also for nonvanishing $\delta\rho_{\phi}$ and $\delta{p}_{\phi}$.

The evolution of $\delta\phi = \phi - \bar{\phi}$ is determined by combining the Klein-Gordon equations for the bubble
 (\ref{kg_bub}) with that of the FRW background (\ref{kg}):
\bea \label{eq:sc_pert_kg} \delta\ddot\phi &=& -3\bar{H}\,\delta\dot\phi + \left(\frac{\dot\delta_m}{1 + \delta_m} + \beta\,\delta\dot\phi\right)(\dot{\bar\phi} + \delta\dot\phi) \nonumber\\
&-& \left(\at{\phi}{U_{,\phi}} - \at{\bar\phi}{U_{,\phi}}\right) + {\beta}\,\delta\rho_m + \frac{\Gamma_{\phi}}{\dot{\bar\phi} + \delta\dot\phi} \eea

Again, we remark that Eqs. (\ref{eq:sc_dens_cont_evol}) and (\ref{eq:sc_pert_kg}) are obtained by applying standard spherical collapse equations (\ref{eq:sc_f2b}) and (\ref{eq:sc_f2bkg}) to coupled quintessence simply
 by adding a coupling in the conservation equations (\ref{rad_bub}) - (\ref{phi_source}). If linearized, Eqs. (\ref{eq:sc_dens_cont_evol}) and (\ref{eq:sc_pert_kg})
read as shown in the left column of Table \ref{tab:sc_rel_comp}. 

In Table \ref{tab:sc_rel_comp} we compare the equations found for $\ddot\delta_m$ and $\delta\ddot\phi$ obtained from standard spherical collapse to Eqs. (\ref{eq:delta_lin_nonnewt}) and (\ref{eq:phi_lin_nonnewt}) obtained
 from the fully relativistic theory in Newtonian gauge and shown in the central column of Table \ref{tab:sc_rel_comp}. 
 In the right column we display the relativistic equations within the Newtonian limit, corresponding to Eqs. (\ref{eq:deltaphi_newt_limit}) and (\ref{deltadotdot_lin}).
Some remarkable problems become evident. 

Note that we have chosen to display the relatistic equations in Newtonian gauge to analyze time derivatives of $\delta\phi$ in the spherical collapse equations, as well as to weigh the importance of different terms in the Klein-Gordon equation when going to small scales. One should keep in mind that at large scales, these equations are gauge dependent.
\begin{table*}[ht]
\begin{tabular}{|p{.3\textwidth}|p{.3\textwidth}|p{.3\textwidth}|}
  \toprule
  \bf{Spherical collapse} & \bf{Relativistic perturbations} & \bf{Newtonian limit}\\ 
  \hline
  \beann \ddot\delta_{m,L} &=& -2\bar{H}\left(\dot\delta_{m,L} + \beta\,\delta\dot\phi\right) \\ 
  && + \frac{1}{2}\sum_{\alpha}(\delta\rho_{\alpha} + 3\delta{p}_{\alpha}) \\
  && - \beta \delta\ddot\phi \eeann &
  \beann \ddot\delta_{m,L} &=& -2\bar{H} \left(\dot\delta_{m,L} + \beta\,\delta\dot\phi\right) \\
  && + (k/a)^2\left(\Phi + \beta\,\delta\phi\right) \\ 
  && + \beta\,\dot\phi\left(\dot\delta_{m,L} + \beta\,\delta\dot\phi\right) \\
  && - 3\dot\Phi\left(2\bar{H} - \beta\,\dot\phi\right) - 3\ddot\Phi \\
  && - \beta\,\delta\ddot\phi \eeann &
  \beann \ddot\delta_{m,L} &=& \left(\beta\,\dot\phi - 2\bar{H}\right)\dot\delta_{m,L} \\
  && + (k/a)^2\left(\Phi + \beta\,\delta\phi\right) \eeann \\  
  (a) & (b) & (c)
  \\
  \hline
  \beann \delta\ddot\phi &=& -3 \bar{H}\delta\dot\phi - U_{,\phi\phi}\,\delta\phi \\
  && + \beta\delta\rho_{m,L} + (\dot\delta_{m,L} + \beta\,\delta\dot\phi)\,\dot\phi \eeann &
  \beann \delta\ddot\phi &=& -3 \bar{H}\delta\dot\phi - U_{,\phi\phi}\,\delta\phi \\
  && + \beta\rho_m\left(\delta_{m,L} + 2\Phi\right) - (k/a)^2\delta\phi \\
  && - 2\Phi U_{,\phi\phi} + 4\dot\phi\dot\Phi \eeann &
  \beann k^2\,\delta\phi &=& a^2\,\beta\,\delta\rho_{m,L} \nonumber \eeann \\
  (d) & (e) & (f)
  \\
  \hline
  \beann \ddot\delta_{m,L} &=& \left(-2\bar{H} - \beta\dot\phi\right)\dot\delta_{m,L} + \bar{H}\beta\,\delta\dot\phi \\ 
  && + \frac{1}{2}\sum_{\alpha}(\delta\rho_{\alpha} + 3\delta{p}_{\alpha}) \\
  && - \beta^2 \delta\rho_{m,L} \\
  && - \beta^2 \delta\dot\phi \dot\phi + U_{,\phi\phi}\beta\,\delta\phi \eeann &
  \beann \ddot\delta_{m,L} &=& \left(-2\bar{H} + \beta\dot\phi \right) \dot\delta_{m,L} + \bar{H} \beta\,\delta\dot\phi \\
  && + (k/a)^2 \Phi \\
  && + 2(k/a)^2 \beta\,\delta\phi - \beta^2 \delta\rho_{m,L} \\ 
  && + \beta^2\,\delta\dot\phi\,\dot\phi + U_{,\phi\phi}\beta\delta\phi \\
  && - 3\dot\Phi\left(2\bar{H} - \beta\,\dot\phi\right) - 3\ddot\Phi \\
  && + 2\beta\Phi U_{,\phi\phi} - 4\beta\dot\phi\dot\Phi \eeann &
  \beann \ddot\delta_{m,L} &=& \left(\beta\,\dot\phi - 2\bar{H}\right)\dot\delta_{m,L} \\
  && + \frac{1}{2}\sum_{\alpha}\delta\rho_{\alpha}\\
  && + \beta^2\delta\rho_{m,L} \eeann \\  
  (g) & (h) & (i)
  \\
  \hline
\end{tabular}
\caption{Comparison between linearized spherical collapse and fully relativistic linear evolution equations. The third row is a combination of the first two rows.}
\label{tab:sc_rel_comp}
\end{table*}

\subsection{Lack of the fifth force in spherical collapse} \label{lack_fifthforce}
As compared to both the Newtonian and relativistic equations, major terms are missing in the standard spherical collapse scenario:

\begin{itemize}
  \item{ Comparing (a) to (b) and (c) in Table \ref{tab:sc_rel_comp}, no term proportional to $\beta \dot{\bar\phi}$ appears.
Depending on the strength of the coupling $\beta$, this term, originating in momentum conservation, can be of great relevance. For $\beta \sim 1$, 
it can significantly alter structure formation when correctly considered in the vectorial velocity equations, as shown in \cite{baldi_etal_2009}. 
For large couplings, as e.g. in a growing neutrino scenario we will discuss later, it is less important, since the cosmon 
$\bar\phi$ is almost constant at late times. Comparing (g) to (h) and (i), one notices a sign reversal in front of a friction-like term. This will also lead to wrong results.}
\item{ Comparing (a) to (b) and (c), as well as (g) to (h), in Table \ref{tab:sc_rel_comp}, terms proportional to $ \beta k^2 \delta\phi$ are absent in the spherical collapse equations; 
  this term is exactly what provides the fifth force. Its absence in (g) leads to a sign reversal of $\beta^2\,\delta\rho_{m,L}$} as compared to (i), thus yielding an incorrect effective Newton's constant.
\item{ Comparing (d) to (e) and (f) in Table \ref{tab:sc_rel_comp}, terms proportional to $ k^2 \delta\phi$ are absent in the spherical collapse equations; 
remarkably, this term does not depend on $\beta$ and is therefore missing even in the uncoupled quintessence scenario}. 
\end{itemize}
The lack of terms proportional to $ \beta k^2 \delta\phi$ leads to a description which does not correspond to the desired coupled quintessence scenario: 
indeed, these are exactly the terms responsible for the fifth force, originating in (\ref{eq:deltaphi_newt_limit}) and leading
 to an effective gravitational force as in Eq.\,(\ref{eq:G_eff}). In other words, we point out that the standard spherical collapse, 
as used for example in \cite{nunes_mota_2006} does not include the main ingredient of coupled quintessence. A fifth attractive force acting between CDM particles and mediated by the cosmon is absent, although densities are indeed coupled to each other as
 in (\ref{mat_bub}) - (\ref{kg_bub}). The reason
 for this can be seen as follows: spherical collapse is by construction based on gravitational dynamics and cannot account
 for other external forces unless appropriately modified. The dynamics in the spherical collapse models are governed by
 the usual Friedmann equations, which are particular formulations of Einstein's field equations. Hence, only 
gravitational forces determine the evolution of the different scale factors and, in turn, of the density contrast. We
 note that, though in the limit of small couplings the difference can be small, for strongly coupled scenarios a completely
 different evolution is obtained. This is simply connected to the fact that for small couplings gravity is still the crucial
 ingredient to fuel the collapse.

\subsection{Inhomogeneity of the scalar field} \label{inhomogeneity}
 The issue of whether the scalar field should be considered to be homogeneous (with the cosmon inside the top hat given by the homogeneous
 background field) or not has also been addressed in literature. In particular, one could try to compare a homogenous scalar field $\phi$
 to an inhomogeneous one by appropriately fixing $\Gamma_\phi$ in Eq.(\ref{kg_bub}) to the expression (\ref{phi_source}) and (\ref{phi_source_inhom})
respectively. This comparison led, for example, \cite{nunes_mota_2006} to find differences
 between the homogenous and inhomogeneous cases. 

The difference found following such procedure is, however, not caused by the fifth force, which, as shown, is not present.
Even without any fifth force or coupling, 
we have noticed in Table \ref{tab:sc_rel_comp}, comparing (d) to (e), that 
terms proportional to $ k^2 \delta\phi$ are absent in the spherical collapse equations.

Evaluating the clustering $\delta \phi$ using a spherical collapse scenario as given by (d) in Table \ref{tab:sc_rel_comp},
leads to effects which do not correspond to the relativistic behavior.
In fact, in absence of the term  $- k^2 \delta \phi$, spherical collapse overestimates the time dependence of the scalar field perturbations 
as soon as $\delta \phi$ is assumed to be different from zero: if, for example, $\delta \phi_\text{in} > 0$ initially, the $\delta\ddot\phi$ obtained
 from the bottom left equation (d) in Table \ref{tab:sc_rel_comp}, is bigger than it would be if the term
 $- k^2 \delta \phi$ appearing in the relativistic equations was actually present. Hence, within spherical collapse, 
all time derivatives of the cosmon are overestimated and, as 
a consequence, $\ddot\delta_m$ is incorrectly reduced. 

We remark that this reasoning also applies to ordinary, uncoupled quintessence. 
Here the question of \mbox{(in)homogeneities} in the scalar field was addressed in various works, 
e.g. \cite{mota_vandebruck_2004, maor_lahav_2005, wang_2005, dutta_maor_2006, mota_etal_2007}.  
Also in this case, one may gain some insight by considering the equations in Table \ref{tab:sc_rel_comp}, now for $\beta = 0$. 
In the relativistic description, scalar field perturbations will decay due to the presence of the term $-(k/a)^2\delta\phi$, until 
the latter may eventually be countered by gravitational contributions. In the spherical collapse, however, this decay is only driven by 
the scalar mass term $-U_{, \phi \phi} \delta \phi$. For a light scalar field and/or sufficiently small scales, this term
is smaller than the missing term $-(k/a)^2\delta\phi$ and therefore scalar field inhomogeneities are incorrectly overestimated. 
The error may be substantially reduced if heavy scalar fields are considered. 
In this case, the presence of the large mass term $V''(\phi)\delta\phi$ in the perturbed Klein-Gordon equation 
may make up for the lack of spatial gradients. 

In conclusion, we have shown that applying the spherical collapse equations to coupled quintessence by merely modifying 
the conservation equations can lead to results which do not correspond to the wanted cosmological scenario: this procedure 
in fact not describe the nonlinear evolution in CQ. 
It is, however, possible to amend the above model to properly include the fifth force whenever a coupling is present.
In order to illustrate and justify that, we consider the nonlinear hydrodynamical evolution equations within coupled quintessence scenarios.

\section{Hydrodynamical spherical collapse: a consistent approach to coupled quintessence}\label{sec:spher_coll_coupled_cor}

In uncoupled, purely gravitational cosmologies the spherical collapse can be derived from the hydrodynamical Navier-Stokes equations. This is a consequence of the fact that the Friedmann equations in presence of nonrelativistic components can be derived from Newtonian gravity. We will demonstrate now that this is also possible in the presence of external forces, basing our analysis on an idea first developed in \cite{wintergerst_2009}.

In order to derive the correct formulation in coupled quintessence, we consider the full nonlinear evolution equations in
 coupled cosmologies within the Newtonian limit:
\bea \label{eq:sph_com_ns1} \dot\delta_m &=& -{\bf v}_m\,\nabla\delta_m - (1 + \delta_m)\,\nabla\cdot{\bf v}_m \\
\dot{\bf v}_m &=& -(2{\bar H} - \beta\,\dot{\bar\phi})\,{\bf v}_m - ({\bf v}_m\,\nabla){\bf v}_m \nonumber \\
\label{eq:sph_com_ns2} && - a^{-2}\,\nabla(\Phi - \beta\,\delta\phi) \\
\label{eq:sph_com_poisson} \Delta\delta\phi &=& -\beta\,a^2\,\delta\rho_m \\
\label{eq:sph_com_grav_pot} \Delta\Phi &=& -\frac{a^2}{2}\,\sum_{\alpha} \delta\rho_{\alpha} \eea
These equations can be derived both from the nonrelativistic Navier-Stokes equations and from the Bianchi identities in the appropriate limit in presence of an external source \cite{kodama_sasaki_1984}. 
\be
\label{eq:ps_cons}\nabla_{\gamma}T_{\mu}^{\gamma} = Q_{\mu} = -\beta T_{\gamma}^{\gamma}  \partial_{\mu}\phi \vv
\ee
where $T^{\gamma}_{\mu}$ is the stress energy tensor of the dark matter fluid. 
They are valid for arbitrary quintessence potentials as long as the scalar field is sufficiently light, i.e. $m_\phi^2 \delta\phi = V''(\phi)\delta\phi \ll \Delta\delta\phi$ for the scales under consideration. For a more detailed discussion of the equations, see \cite{wintergerst_etal_2009, wintergerst_2009}.
We are working in comoving spatial coordinates $\bf x$ and cosmic time $t$. The sign in Eq. (\ref{eq:sph_com_grav_pot}) was chosen to match Eq. (\ref{eq:laplace_grav_pot}).
Note that ${\bf v}_m$ is the comoving velocity, related to the peculiar velocities by ${\bf v}_m = {\bf v}_{pec}/a$. The sum
 in Eq.(\ref{eq:sph_com_grav_pot}) is to be taken over all clustering components; as an important consequence
 of the Newtonian limit, the cosmon is explicitly excluded.

In order to obtain a correct description of the spherical collapse model, we are interested in the evolution of a top hat,
 spherically symmetric around ${\bf x} = 0$. We note that the below derivation is not limited to a top hat but holds for the amplitude at ${\bf x} = 0$ for generic spherically symmetric profiles. From simple symmetric arguments we may infer
\be \label{eq:zero_con} \at{{\bf x} = 0}{\nabla\delta_m} = {\bf v_m}(0,t) = 0 \vv \ee
which changes \eqref{eq:sph_com_ns1} to
\be \label{eq:new_ns1} \at{ {\bf x} = 0}{\dot\delta_m} = - \at{ {\bf x} = 0}{\left[(1 + \delta_m)\,\nabla\cdot{\bf v}_m\right]} \pp \ee
We now want to relate Eqs. \eqref{eq:sph_com_ns1}-\eqref{eq:sph_com_grav_pot} to the spherical infall: it is therefore useful to combine \eqref{eq:sph_com_ns1} and \eqref{eq:sph_com_ns2} to give a second order equation for $\delta_m$, taken at ${\bf x} = 0$
\be \at{{\bf x} = 0}{\ddot\delta_m} = \at{ {\bf x} = 0}{\frac{\dot\delta_m^2}{1+\delta_m}} - \at{ {\bf x} = 0}{\left[(1+\delta_m)\,\nabla\cdot\dot{\bf v}_m\right]} \vv \ee
where we have used \eqref{eq:zero_con} and \eqref{eq:new_ns1}. Inserting the divergence of \eqref{eq:sph_com_ns2} yields
\bea \label{eq:delta_bef_lim} \at{{\bf x} = 0}{\ddot\delta_m} &=& \at{ {\bf x} = 0}{-(2{\bar H}-\beta\,\dot{\bar\phi})\,\dot\delta_m} \nonumber \\ 
&+&  \at{ {\bf x} = 0}{\left[\frac{\dot\delta_m^2}{1+\delta_m} + \frac{1 + \delta_m}{a^2}\,\Delta\Phi_{\text{eff}}\right]} \nonumber \\
&+& \at{ {\bf x} = 0}{(1+\delta_m)\nabla({\bf v}_m\,\nabla){\bf v}_m} \pp \eea
Note that $\Phi_{\text{eff}}$ is defined as in \eqref{eq:eff_grav_pot} and obeys the Laplace equation \eqref{eq:laplace_grav_pot}, 
as can be seen by combining \eqref{eq:sph_com_poisson} and \eqref{eq:sph_com_grav_pot}. 
The first three terms in \eqref{eq:delta_bef_lim} can be evaluated straightforwardly at ${\bf x} = 0$. To rewrite the last term we use the identity 
\bea \label{eq:double_nabla_v} \at{ {\bf x} = 0}{\nabla({\bf v}_m\,\nabla){\bf v}_m} &=& \frac{1}{3}\at{ {\bf x} = 0}{(\nabla\cdot{\bf v}_m)^2} \\
&=& \frac{1}{3}\at{ {\bf x} = 0}{\frac{\dot\delta_m^2}{(1+\delta_m)^2}} \nonumber \eea
which holds for spherically symmetric situations and is rederived in the Appendix.
Inserting this into expression \eqref{eq:convec_bef_lim} and subsequently into \eqref{eq:delta_bef_lim} yields the final expression for the evolution of the top hat density amplitude (writing $\delta$ instead of $\at{ {\bf x} = 0}{\delta}$)
\mybox{
\bea \ddot\delta_m &=& -(2{\bar H}-\beta\,\dot{\bar\phi})\,\dot\delta_m \nonumber \\
 \label{eq:sph_gen_del} &+& \frac{4}{3}\frac{\dot\delta_m^2}{1 + \delta_m} + \frac{1 + \delta_m}{a^2}\,\Delta\Phi_{\text{eff}} \pp \eea
 }
Linearization leads to: 
\mybox{
\be \label{eq:sph_gen_del_lin} \ddot\delta_{m,L} = -(2{\bar H}-\beta\,\dot{\bar\phi})\,\dot\delta_{m,L} + a^{-2}\,\Delta\Phi_{\text{eff}} \vv \ee
}
which corresponds to the relativistic equation (\ref{deltadotdot_lin}).
Here we recall that the effective gravitational potential, given by (\ref{eq:eff_grav_pot}), follows the modified Poisson equation (\ref{eq:laplace_grav_pot})
which we rewrite here for convenience:
\be \Delta {\Phi_{\text{eff}}} = -\frac{a^2}{2} {\bar\rho}_m \delta_m \left(1+2 \beta^2\right) \, \pp \ee
Equations (\ref{eq:sph_gen_del}) and (\ref{eq:sph_gen_del_lin}) are the two main equations which correctly describe the nonlinear and linear evolution for
a coupled dark energy model. They describe the dynamics of a spherical top hat as it follows 
from relativistic perturbation theory in the Newtonian regime and they can be used, among other things,
 for estimating the extrapolated linear density contrast at collapse $\delta_c$ in the presence of
a fifth force. To our knowledge it is the first time that the second order equations (\ref{eq:sph_gen_del}) and (\ref{eq:sph_gen_del_lin}) 
are presented in this way.

We will now demonstrate that we may easily reformulate Eqs. (\ref{eq:sph_gen_del}) and (\ref{eq:sph_gen_del_lin}) into 
an effective spherical collapse: we can combine them to derive an equation for the radius $r$ which extends Eq.(\ref{eq:sc_f2b}) to the case 
of coupled dark energy.
To do so, we consider a spherical bubble of radius $r$ containing the CDM overdensity $\delta_{m}$. Particle number conservation yields
\be \label{eq:part_num_cons} 1 + \delta_{n,m} = (1 + \delta_{n,m,\text{in}})\,\left(\frac{r_\text{in}}{a_\text{in}}\right)^3\,\left(\frac{a}{r}\right)^3 \vv \ee
where $n$ is the number density of CDM particles and $\delta_n \equiv \delta{n}/n$.

We demand the scale factors $r$ and $a$ to be equal initially, i.e. $a_\text{in} = r_\text{in}$. Further, we assume that the mass of CDM particles is the same inside the bubble and in the background. Note that this is not a limitation, but merely a prescription that we have employed in order to obtain an equation for the scale factor $r$ in a form which is analogous to the original Friedmann equation (\ref{eq:sc_f2b}). We obtain
\be \label{eq:delta_cdm} 1 + \delta_m = (1 + \delta_{m,\text{in}})\,\left(\frac{a}{r}\right)^3 \pp \ee
The first and second time derivatives of $\delta_m$ then read
\bea \label{eq:sc_cont_dot} \dot\delta_m &=& 3 \left(1 + \delta_m\right) \left(\frac{\dot a}{a} - \frac{\dot r}{r}\right) \vv \\
\ddot\delta_m &=& 3\left(1 + \delta_m\right)\,\left(\frac{\ddot a}{a} - \frac{\ddot r}{r} + \left(\frac{\dot r}{r}\right)^2 - \left(\frac{\dot a}{a}\right)^2\right) \nonumber \\
\label{eq:sc_cont_ddot} && + \frac{\dot\delta_m^2}{1 + \delta_m} \vv \eea
which we can combine appropriately to yield
\be \label{eq:sc_cont_ddot_final} \ddot\delta_m = -2{\bar H}\,\dot\delta_m + \frac{4}{3}\,\frac{{\dot\delta_m}^2}{1+\delta_m} + 3(1+\delta_m)\left(\frac{\ddot a}{a} - \frac{\ddot r}{r}\right) \pp \ee
Comparison to \eqref{eq:sph_gen_del} and insertion of the background Friedmann equation \eqref{eq:sc_f1b} gives the evolution equation for the bubble radius
\mybox{
\bea \frac{\ddot{r}}{r} &=& -\beta\,\dot{\bar\phi}\left(\bar H - \frac{\dot{r}}{r}\right) - \frac{1}{6} \sum_{\alpha} \left[{\bar\rho}_{\alpha}(1 + 3 {\bar w}_{\alpha})\right] \nonumber \\
\label{eq:imp_sc_f2b} &-& \frac{1}{3}\,\beta^2\,\delta\rho_m \pp \eea
}
Equation (\ref{eq:imp_sc_f2b}), equivalent to the one used in \cite{mainini_bonometto_2006}, describes the general evolution of
the radius of a spherical overdense region within coupled quintessence. 
Comparing with the Friedmann equation (\ref{eq:sc_f2b}) we notice the presence of two additional terms: a friction term 
and the coupling term $\beta^2\,\delta\rho_m$; the latter is precisely the term responsible for the additional attractive fifth force.
Note that the ``friction'' term is velocity dependent and its effects on collapse depend, more realistically, on the
direction of the velocity \cite{baldi_etal_2009}, information which is not contained within a \emph{spherical} collapse picture.

We conclude that one may indeed apply the spherical collapse model to coupled dark energy scenarios. 
However, it is crucial to include the additional force term in the equations. 

Note that the outlined procedure can easily be generalized to include uncoupled components, for example baryons. 
In this case, the corresponding evolution equation for $\delta_b$, will be fed by $\Phi_{\text{eff}} = \Phi$. 
This yields an evolution equation for the uncoupled scale factor $r_{uc}$ that is equivalent to the regular Friedmann equation (\ref{eq:sc_f2b}).

\subsection{Methods and initial conditions}\label{sec:ini_conds}

To provide maximum stability and to rule out a dependence on initial conditions, we directly integrate Eqs. (\ref{eq:sph_gen_del}) and (\ref{eq:sph_gen_del_lin}) for the nonlinear and linear density contrasts, together with the corresponding background equations and the Klein-Gordon equation (\ref{kg}) for the scalar field. The radial parameter $r(z)$ may equivalently be obtained by integrating Eq.\,(\ref{eq:imp_sc_f2b}) or by directly applying the relation (\ref{eq:delta_cdm}). The following initial conditions at the initial redshift $z_\text{in}$ were imposed:
\begin{itemize}
  \item $\delta_{m,\text{in}} = \delta_{m,L,\text{in}}$
  \item $\dot\delta_{m,L,\text{in}} = 3(1 + \delta_{m,L,\text{in}})(\bar{H}_\text{in} - {H}_\text{in}) = 0$, as initially the Hubble functions
 of background and overdensity evaluate to the same value.
\end{itemize}
The value of the extrapolated linear density contrast at collapse $\delta_c$ can be obtained by stopping the evolution of Eq.(\ref{eq:sph_gen_del_lin}) when $\delta_{m}$ as obtained
 from (\ref{eq:sph_gen_del}) goes to infinity, i.e. the overdensity collapses. If we then vary the initial conditions,
 leading to different collapse redshifts $z_c$, we arrive at a redshift dependent expression for this critical
 density, $\delta_c = \delta_c(z_c)$. Equivalently, one may vary the initial redshift $z_\text{in}$, keeping $\delta_{m,\text{in}}$ fixed. 
To be sure of starting the integration when densities are still linear, we find that it is necessary to work in a range of initial overdensities with
 $\delta_{m,\text{in}} < 10^{-3}$. 

\subsection{Results}\label{sec:coupled_results}

\begin{figure}[ht]
\begin{center}
\includegraphics[width=85mm,angle=-0.]{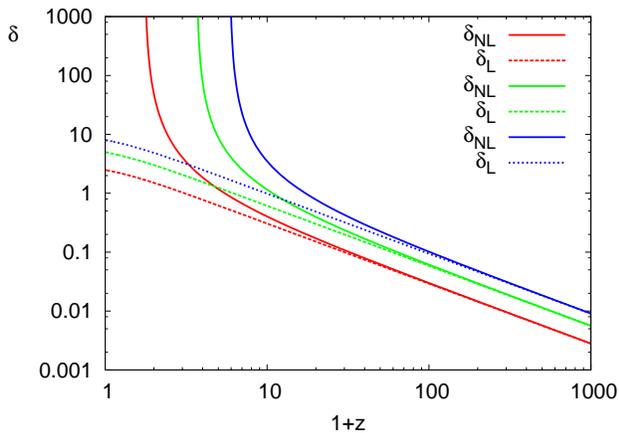}
\end{center}
\caption{CDM linear and nonlinear perturbations for different initial conditions.}
\label{fig:deltas_ccdm}
\vspace{0.5cm}
\end{figure}

\begin{figure}[ht]
\begin{center}
\includegraphics[width=85mm,angle=0.]{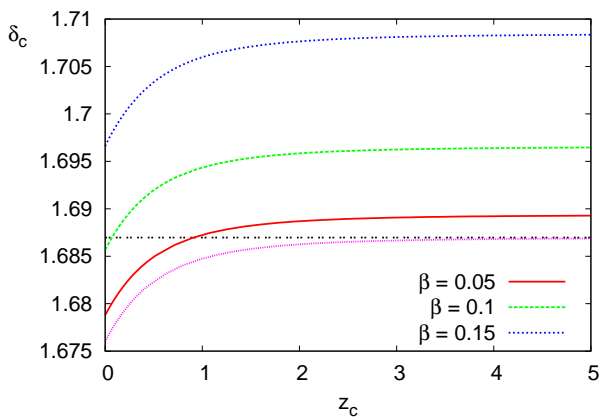}
\end{center}
\caption{Extrapolated linear density contrast at collapse for coupled quintessence models with different coupling strength $\beta$. 
For all plots we use a constant $\alpha = 0.1$. We also depict $\delta_c$ for reference $\Lambda$CDM (dotted, pink) and EdS (double-dashed, black) models.}
\label{fig:delta_c_ccdm}
\vspace{0.5cm}
\end{figure}

We depict the evolution of $\delta_m(z)$ and $\delta_{m,L}(z)$ for different initial redshifts in Fig. \ref{fig:deltas_ccdm}. 
For this plot, we used sample parameters $\alpha = 0.1$ and $\beta = 0.1$. 
 We have also plotted the linear density contrast at collapse $\delta_c(z_c)$ for three coupled quintessence
 models with $\alpha = 0.1$ and $\beta = 0.05$, $0.1$ and $0.15$ in Fig. \ref{fig:delta_c_ccdm}.
We note that these as well as all subsequent results are valid under the hypothesis in which the linear extrapolation traces the nonlinear behavior when a fifth force is present. 

\begin{figure}[ht]
\begin{center}
\includegraphics[width=85mm,angle=0.]{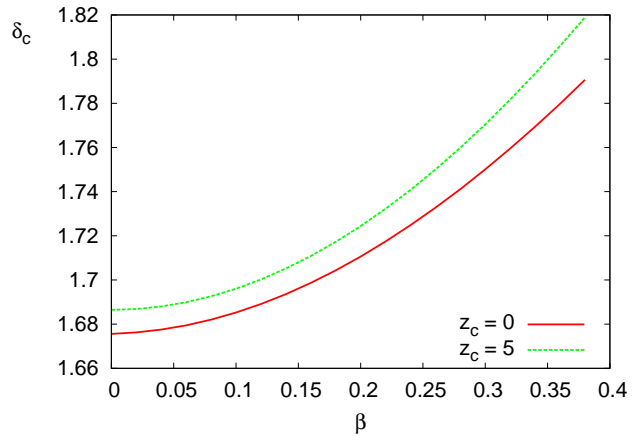}
\end{center}
\caption{Extrapolated linear density contrast at collapse $\delta_c$ for coupled quintessence models as a function of coupling strength $\beta$, evaluated for two different collapse redshifts $z_c = 0$ (solid, red) and $z_c = 5$ (long-dashed, green).}
\label{fig:delta_c_of_beta}
\vspace{0.5cm}
\end{figure}

As opposed to the results found in \cite{nunes_mota_2006}, no oscillations are seen in $\delta_c(z_c)$. Furthermore, 
the effect of the coupling on the extrapolated linear density contrast at collapse is smaller, 
though we observe an increase of $\delta_c$ with increasing coupling strength $\beta$, as depicted in Fig.\ref{fig:delta_c_of_beta} for two collapse redshifts $z_c = 0$ and $z_c = 5$. A coupling $\beta = 0$ corresponds to a $\Lambda$CDM cosmology, hence the observed $\delta_c$ is given by $\delta_c = 1.686$ for z $z_c = 5$ and by the accordingly reduced value for $z_c = 0$. An increase of $\beta$ results in an increase of $\delta_c$ for both redshifts. The reason for this increase is quite simple. In Eqs.(\ref{eq:sph_gen_del}) and (\ref{eq:sph_gen_del_lin}) two terms lead to an enhanced growth: the fifth force term in the effective potential and the reduction of the damping $-\beta\dot{\bar\phi}$. In the linear equation, they are always of comparable strength. In Eq.(\ref{eq:sph_gen_del}), however, the damping will be negligible once $\delta_m \sim 1$ as it only enters the equation linearly. The enhancement of growth is then weaker than in the linear equation and $\delta_c$ grows with increasing $\beta$.

For small $\beta \lsim 0.4$, $\delta_c(\beta)$ at $z_c \geq 5$ is well described by a simple quadratic fitting formula,
\be \delta_c(\beta) = 1.686(1 + a\beta^2)\,,a = 0.556 \pp \ee
For larger $\beta \leq 1$ a fit requires an additional correction and reads
\be \delta_c(\beta) = 1.686(1 + a\beta^2 - b\beta^4),\,a = 0.556,\,b = 0.107 \pp \ee

It is worth noting that our values of $\delta_c$ were obtained under the assumption that baryonic contributions may be neglected, in order to be able to relate the results to the simple Einstein de Sitter scenario. Indeed, a numerical analysis under inclusion of a baryonic component shows a quite significant increase of the critical density contrast $\delta_c$, leading to values close to those found in \cite{mainini_bonometto_2006}.

Also note that we have limited our analysis to the critical density contrast at collapse. Other works \cite{angrick_bartelmann_2010} have rather focused on the respective quantities at virialization. Since there are no fundamental differences, these may as easily be evaluated within our formalism. 

\section{Growing neutrinos}\label{sec:nu_results}

Another interesting framework, analogous to coupled quintessence, in which a fifth force is present, is the growing neutrinos scenario \cite{amendola_etal_2007, wetterich_2007}. Here, relic neutrinos obtain a growing, cosmon dependent mass, implemented by a large, negative coupling $\beta$. In this context, dark energy domination and the late acceleration of the Universe can be naturally explained by relating it to a ``trigger event'', the recent transition of neutrinos to the nonrelativistic regime.

\subsection{Cosmological model}

As neutrinos have been relativistic particles through most of the history of the Universe, Eqs. (\ref{kg}), (\ref{cons_phi}) and (\ref{cons_gr}) are appropriately altered to include neutrino pressure
\bea \label{eq:dens_phi_gn} \dot{\bar\rho}_\phi &=& -3\bar H(1+{\bar w}_\phi){\bar\rho}_\phi + \beta\,\dot{\bar\phi}\,(1 - 3 {\bar w}_\nu){\bar\rho}_\nu \\
\label{eq:dens_nu_gn} \dot{\bar\rho}_\nu &=& -3\bar H(1+{\bar w}_\nu){\bar\rho}_\nu - \beta\,\dot{\bar\phi}\,(1-3 {\bar w}_\nu){\bar\rho}_\nu \\
\label{eq:kg_gn} \ddot{\bar\phi} &=& -3\bar H\dot{\bar\phi} - \frac{dU}{d{\bar\phi}} + \beta(1-3{\bar w}_\nu){\bar\rho}_\nu \pp \eea
As opposed to the models of coupled CDM discussed above, the constant $\beta$ is now negative and its modulus much larger than one. Bounds for the couplings $\alpha$ and $\beta$ have been discussed in \cite{doran_etal_2007, amendola_etal_2007}. For the following analysis we choose $\alpha = 10$ and several values for the coupling $\beta = -52$, $-112$ and $-560$. Note that the couplings may be related to the present neutrino mass via \be m_{\nu}(t_0) = -\frac{\alpha}{\beta}\,\Omega_\phi(t_0)\,16\text{ eV} \vv \ee where $\Omega_\phi$ is the dark energy density fraction today.

\begin{figure}[ht]
\begin{center}
\includegraphics[width=85mm,angle=0.]{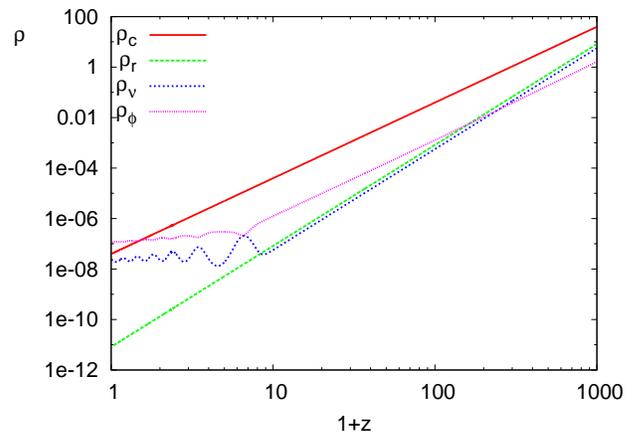}
\end{center}
\caption{Energy densities of neutrinos (solid, red), cold dark matter (long dashed, green), dark energy (dot-dashed, blue) and photons (short dashed, black) are plotted vs redshift. We use a sample model with constant $\beta = -52$, $\alpha = 10$ and a large average neutrino mass $m_\nu = 2.11$ eV.}
\label{fig:gn_bkg}
\vspace{0.5cm}
\end{figure}

A numerical integration of (\ref{eq:dens_phi_gn}) - (\ref{eq:kg_gn}) and the appropriate equations for radiation and CDM leads to the background evolution depicted in Fig. \ref{fig:gn_bkg}. While the cosmon is on the matter (radiation) attractor at early times, the transition of neutrinos to the nonrelativistic regime almost stops the evolution of the cosmon. The dark energy density is able to overcome all other components and dominates the Universe at $t = t_0$. As the kinetic contribution to the cosmon energy density is greatly reduced, the latter is dominated by the potential $V({\bar\phi})$, successfully mimicking the behavior of a cosmological constant.

\subsection{Spherical collapse and growing neutrino quintessence}

We have applied our method to the case of growing neutrino quintessence. To our knowledge this is the first time that spherical collapse is performed on this class of models. 

Because of the strong cosmon-mediated attractive force between neutrinos, bound neutrino structures may form within these models \cite{brouzakis_etal_2007}. It was shown in \cite{mota_etal_2008} that their formation will only start after neutrinos become nonrelativistic. A nonlinear treatment of the evolution of neutrino densities is thus only required for very late times, and one may safely neglect neutrino pressure as compared to their density, which substantially simplifies the scenario. All calculations of the previous sections are thus equally valid for the growing neutrinos scenario; we can straightforwardly apply the evolution equations (\ref{eq:sph_gen_del}) and (\ref{eq:sph_gen_del_lin}) for the nonlinear and linear neutrino density contrast.

\begin{figure}[ht]
\begin{center}
\includegraphics[width=85mm,angle=0.]{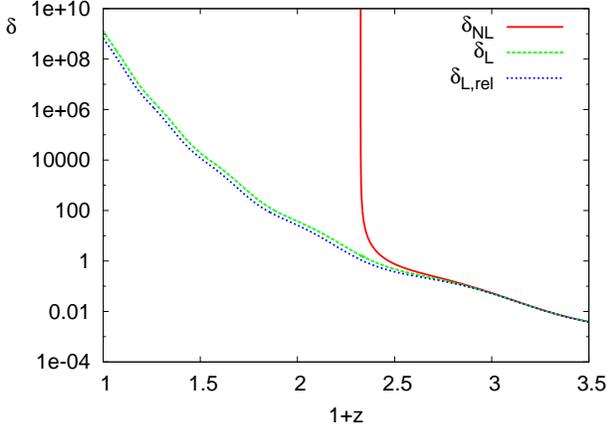}
\end{center}
\caption{Evolution of neutrino nonlinear (solid, red) and linear (long-dashed, green) density contrast. For comparison, we have also included the relativistic linear density contrast including pressure terms (short-dashed, blue).}
\label{fig:deltas_cnu}
\vspace{0.5cm}
\end{figure}

In Fig.\ref{fig:deltas_cnu} we plot the evolution of the nonlinear density contrast as obtained from numerically 
solving Eq.\,(\ref{eq:sph_gen_del}) for a model with $\beta = -52$. The linear density contrast, solution of Eq.\,(\ref{eq:sph_gen_del_lin}), is also shown. 
For comparison, we have also included the linear density contrast resulting from the full relativistic equations as given
in \cite{mota_etal_2008}. The results of the linearized spherical collapse and the relativistic theory can be seen to agree remarkably well. Comparison with the full hydrodynamic results from Ref.\cite{wintergerst_etal_2009} also yields a one-to-one agreement, as expected since the latter is the basis for the present work. 
 The slight deviation around redshift $z \sim 1.5$ can be accounted for by a short recuperation of neutrino
 pressure at this time. However, no significant impact on the extrapolated linear density contrast $\delta_c$ was found.

\begin{figure}[ht]
\begin{center}
\includegraphics[width=85mm,angle=0.]{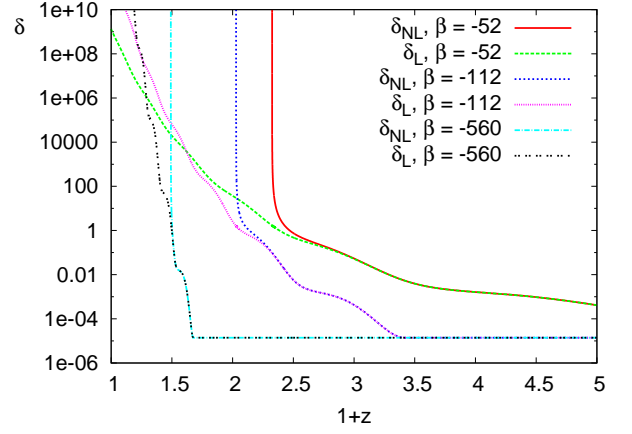}
\end{center}
\caption{Evolution of neutrino nonlinear / linear density contrast for $\alpha = 10$ and $\beta = -52$ (solid, red / long-dashed, green), $\beta = -112$ (short-dashed, blue / dotted, pink) and $\beta = -560$ (dot-dashed, light blue / double-dashed, black). The chosen couplings correspond to a present average neutrino mass of $m_\nu(t_0) = 2.1$ eV, $1$ eV and $0.2$ eV, respectively.}
\label{fig:deltas_cnu_3cpl}
\vspace{0.5cm}
\end{figure}

In order to illustrate the dependence of the growth of the overdensity on the coupling $\beta$, we show the evolution
 of $\delta_\nu(z)$ and $\delta_{\nu,L}(z)$ in Fig. \ref{fig:deltas_cnu_3cpl} for the three given
 couplings $\beta = -52,\,-112,\,-560$ and $\alpha = 10$. Given a fixed self-interaction $\alpha$,
 a larger $\beta$ corresponds to a smaller present neutrino mass; as a consequence, neutrinos become nonrelativistic at smaller redshifts $z_\text{NR}$.
 In the relativistic regime, no growth of neutrino perturbations is observed in the linear regime. To comply with this, we only start the integration of the spherical collapse equations once the transition to the nonrelativistic regime is observed at redshift $z_\text{NR}(\beta)$.
 It can be observed in Fig.\ref{fig:deltas_cnu_3cpl} that a higher $\beta$ leads to a strongly enhanced growth of the density contrast. On the other hand, because of the later transition to the nonrelativistic regime, perturbations start to grow at much lower redshifts.

\begin{figure}[ht]
\begin{center}
\includegraphics[width=85mm,angle=0.]{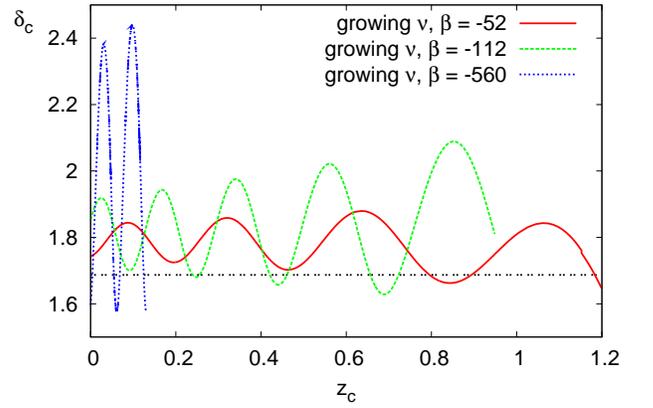}
\end{center}
\caption{Extrapolated linear density contrast at collapse $\delta_c$ vs. collapse redshift $z_c$ for growing neutrinos with $\beta = -52$ (solid, red), $\beta = -112$ (long-dashed, green) and $\beta = -560$ (short-dashed, blue). A reference EdS model (double-dashed. black) is also shown.}
\label{fig:delta_c_cnu}
\vspace{0.5cm}
\end{figure}

The extrapolated linear density at collapse $\delta_c$ for growing neutrino quintessence reflects in all respects 
the characteristic features of this model and result in a $\delta_c$ which looks quite different from standard dark energy cosmologies.
We have plotted the dependence of $\delta_c$ on the collapse redshift $z_c$ in Fig.\ref{fig:delta_c_cnu} for all three couplings. 

The oscillations seen are the result of the oscillations of the neutrino mass caused by the coupling to the scalar field: the latter 
has characteristic oscillations as it approaches the minimum of the effective potential in which it rolls, given by a combination of
the self-interaction potential $U(\phi)$ and the coupling contribution $\beta(1-3{\bar w}_\nu){\bar\rho}_\nu$. 
Furthermore, due to the strong coupling $\beta$, the average
 value of $\delta_c$ is found to be substantially higher than $1.686$.
Such an effect can have a strong impact on structure formation and $\delta_c$ can then be used within a Press-Schechter formalism.

For the strongly coupled models, corresponding to a low present day neutrino mass $m_\nu(t_0)$, the critical density at collapse is only available for $z_c \lsim 0.2$, $1$ for $\beta = -560$, $-112$, respectively. This is again a reflection of the late transition to the nonrelativistic regime.

A full nonlinear investigation of single lumps within growing neutrino quintessence was performed in \cite{wintergerst_etal_2009}.

\section{Early dark energy}\label{sec:early_results}
\subsection{Cosmological model}
A convenient way to parametrize the presence of a nonnegligible homogenous dark energy component at early times (from now on labeled as EDE) was presented in \cite{wetterich_2004}. Here, the dark energy density is \be {\bar\rho}_{\text{DE}}(z) = {\bar\rho}_{\text{DE},0} \left(1+z\right)^{3(1+\bar{w}_h(z))} \vv \ee with \be {\bar\rho}_{\text{DE},0} = {\bar\rho}_{crit,0}\,\Omega_{\text{DE},0} = 3{\bar H_0}^2\left(1-\Omega_{m,0}\right) \ee and the equations of state parametrized by:
\be \bar{w}_h (z) = \frac{{\bar w}_0}{1+b \ln{(1+z)}} \vv \ee
where $b$ is a constant related to the amount of dark energy present at early times \be b = - \frac{3 {\bar w}_0}{\ln{\frac{1-\Omega_{\text{DE},e}}{\Omega_{\text{DE},e}}} + \ln{\frac{1-\Omega_{m,0}}{\Omega_{m,0}}}} \pp \ee
Here the subscripts ``$0$'' and ``$e$'' refer to quantities calculated today or early times, respectively.
Dark energy pressure will be given by $p_{\text{DE}}(z) = \bar{w}_{h}(z)\,{\bar\rho}_{\text{DE}}(z)$.
If we specify the spherical collapse equations for this case, the nonlinear evolution of the density contrast follows the evolution equations (\ref{eq:sph_gen_del}) and (\ref{eq:sph_gen_del_lin}) without the terms related to the coupling: 
\bea \ddot\delta_m &=& -2{\bar H} \dot\delta_m + \frac{4}{3} \frac{{\dot\delta_m}^2}{1+\delta_m} + \frac{1}{2} \delta_m (1 + \delta_m) {\bar\rho}_m \vv \\
\ddot\delta_{m,L} &=& -2{\bar H} \dot\delta_{m,L} + \frac{1}{2} \delta_{m,L}\,{\bar\rho}_m \pp \eea
As before, we assume relativistic components to remain homogenous.
\subsection{Spherical collapse and EDE}
In the following we present our results for two models of early dark energy, namely model I and II from \cite{bartelmann_etal_2006}. Model I is given by the set of parameters
\be \label{eq:bdw_I} \Omega_{m,0} = 0.332\,,\,\,\,w_0 = -0.93\,,\,\,\,\Omega_{\text{DE},e} = 2\cdot10^{-4} \vv \ee
whereas model II is parametrized by
\be \label{eq:bdw_II} \Omega_{m,0} = 0.314\,,\,\,\,w_0 = -0.99\,,\,\,\,\Omega_{\text{DE},e} = 8\cdot10^{-4} \pp \ee

\begin{figure}[ht]
\begin{center}
\includegraphics[width=85mm,angle=0.]{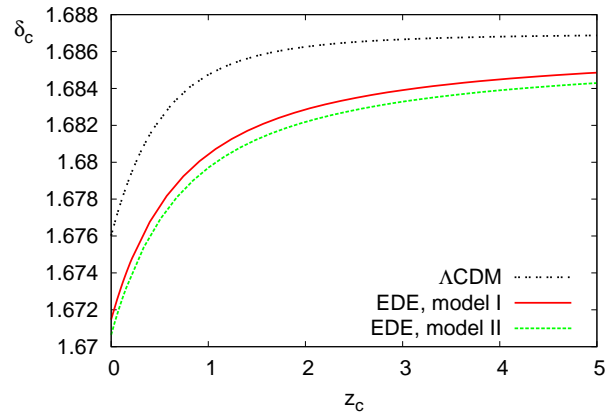}
\end{center}
\caption{Extrapolated linear density contrast at collapse $\delta_c$ vs. collapse redshift $z_c$ for EDE models I (solid, red) and II (long-dashed, green), as well as $\Lambda$CDM (double-dashed, black).}
\label{fig:delta_c_ede}
\vspace{0.5cm}
\end{figure}

Our results for $\delta_c$ in both EDE models are plotted in Fig.\,\ref{fig:delta_c_ede}, together with $\delta_c$ in $\Lambda$CDM. Coherent with the results of \cite{francis_etal_2008}, we find a suppression of $\delta_c$ as compared to $\Lambda$CDM that is much lower than in the original paper \cite{bartelmann_etal_2006}. More precisely, while the latter found $\delta_c(z_c = 5) \sim 1.62$ for model I, corresponding to a relative change of $\sim 4\%$, we obtain $\delta_c(z_c = 5) \sim 1.685$ ($\sim 5\cdot10^{-2} \%$).

\section{Conclusions}
Spherical collapse is a semi analytical method often used to estimate the nonlinear evolution of structures 
without reverting to complex numerical methods like N-body simulations. 

After reviewing its application to standard cosmologies, 
we have considered the case of coupled dark energy cosmologies, in which a fifth force other than gravity modifies the collapse. 

We have shown that the inclusion of the fifth force within the spherical collapse picture deserves particular caution.
As spherical collapse is intrinsically based on gravitational attraction via the Friedmann equations, it does not account
 for other external forces unless it is suitably modified.

We have presented a detailed comparison between the linearized standard spherical collapse picture
 and the linear relativistic equations, whose results are summarized in Table \ref{tab:sc_rel_comp}. 
Applying standard spherical collapse equations to coupled dark energy by adding a coupling in the conservation equations
is insufficient to describe the coupled dark energy scenario, as the fifth force is still missing entirely from the evolution of 
the density contrast $\delta$. Results in Table \ref{tab:sc_rel_comp}
also show that a standard treatment of spherical collapse may lead to problems even in the uncoupled case,  
whenever the scalar field is treated as inhomogeneous. 

We have illustrated in detail how a modification of the spherical collapse picture which correctly accounts for the presence of a fifth force
 is still possible.  
We have derived the set of second order differential equations for the density contrast from the fully nonlinear Navier-Stokes equations. We have then shown how $\delta_c$ can be evaluate directly from these equations and how the spherical collapse formalism can be reformulated starting from them. 
Most importantly, we have further checked that our results match the numerical resolution
of the nonlinear hydrodynamical equations performed as described in \cite{wintergerst_etal_2009, wintergerst_2009}.

We have applied our procedure to coupled quintessence scenarios, evaluating the extrapolated linear density at collapse
for this class of cosmologies and showing how it depends on the coupling $\beta$. 
Furthermore, we have for the first time applied the spherical collapse to the case of growing neutrino quintessence, where neutrinos feel a fifth force interaction
 that can lead them to cluster at very large scales. In this case, we demonstrate how the extrapolated linear density 
at collapse shows a characteristic oscillating behavior, different from standard dark energy models. In future work, this result could be used within a Press-Schechter \cite{press_schechter_1974} formalism to estimate neutrino halo
 mass distributions.
We have further commented on the choice of initial conditions, whose choice has to be made with careful attention when dealing with
the extrapolated linear density at collapse. 
Finally we have used our approach to verify results found in \cite{francis_etal_2008} on spherical collapse and early dark energy (EDE).

\section*{ACKNOWLEDGEMENTS}
We are very grateful to David F. Mota and Christof Wetterich for precious help for this work.
We also acknowledge Matthias Bartelmann, Mischa Gerstenlauer and Francesco Pace for useful discussions. The work of N.W. is supported by the Humboldt Foundation.

\section*{APPENDIX}
In order to derive the identity (\ref{eq:double_nabla_v}, it is convenient to express $\nabla({\bf v}_m\,\nabla){\bf v}_m$ in spherical coordinates. 
Symmetry implies ${\bf v}_m = {\rm v}_m\,{\bf e}_r$ and therefore
\be \nabla({\bf v}_m\,\nabla){\bf v}_m = \left[(\nabla\cdot{\bf v}_m)^2 - \frac{2}{r}\,{\rm v}_m\,\nabla\cdot{\bf v}_m + {\rm v}_m\,\partial_r^2{\rm v}_m\right] \vv \ee
which we may evaluate at ${\bf x} = 0$ making use of \eqref{eq:zero_con} and \eqref{eq:delta_bef_lim}:
\be \label{eq:convec_bef_lim} \at{ {\bf x}=0}{\nabla({\bf v}_m\,\nabla){\bf v}_m} = \at{ {\bf x} = 0}{(\nabla\cdot{\bf v}_m)^2} - 2\nabla\cdot{\bf v}_m \mathop {\lim }\limits_{r \to 0}\frac{{\rm v}_m}{r} \pp \ee
To evaluate the limit, we consider a scalar function defined as
\be \label{eq:sph_mass} \xi(r,t) := 4\pi\int_0^r r'^2 \delta_m(r',t) dr' \pp \ee
We differentiate it in time and insert (\ref{eq:sph_com_ns1}) in spherical coordinates, using $\nabla\cdot\left[(1+\delta_m){\bf v}_m\right] = \frac{1}{r^2}\partial_r\left[r^2(1+\delta_m){\rm v}_m\right]$. Integration yields the following expression for the velocity:
\be {\rm v}_m(r,t) = -\frac{\dot\xi}{4\pi r^2(1 + \delta_m)} \ee
Therefore 
\be \mathop {\lim}\limits_{r \to 0}\left(\frac{{\rm v}_m}{r}\right) = - \mathop {\lim }\limits_{r \to 0}\left(\frac{\dot\xi}{4\pi r^3(1+\delta_m)}\right) \ee
Expanding $\delta(r',t)$ in (\ref{eq:sph_mass}) around $r' = 0$ enables us to rewrite $\xi$ as
\beann \xi(r,t) &=& 4\pi\int_0^r r'^2\,\left(\delta_m(0,t) + O(r')\right)\,dr' \\
&=& \frac{4\pi}{3}\,r^3\,\delta_m(0,t) + O(r^4) \eeann
and thus
\be \mathop {\lim }\limits_{r \to 0}\left({\frac{\dot\xi}{4\pi r^3 (1+\delta_m)}}\right) = -\frac{1}{3}\at{ {\bf x}=0}{\nabla\cdot{\bf v}_m} \pp \ee
We obtain
\be \at{ {\bf x}=0}{\nabla({\bf v}_m\,\nabla){\bf v}_m} = \frac{1}{3}\at{ {\bf x}=0}{(\nabla\cdot{\bf v}_m)^2} \ee

\end{document}